\documentclass[prd,amsmath,amssymb,aps,twocolumn,preprintnumbers,showpacs,floatfix]{revtex4-1}

\usepackage{color,graphicx} 
\usepackage{float}

\begin{document}

\preprint{DESY 13-107, DO-TH 13/15}
\title{Hadroproduction of 
$\Upsilon(nS)$ above $B\bar B$ Thresholds
and Implications for the $Y_b(10890)$
}

\author{Ahmed Ali}
\affiliation{ Deutsches Elektronen-Synchrotron DESY, Hamburg 22607, Germany }
\author{Christian Hambrock}
\affiliation{Institut f\"ur Physik, Technische Universit\"at Dortmund,
Dortmund 44221, Germany }
\author{Wei Wang}
\affiliation{Helmholtz-Institut f\"ur Strahlen- und Kernphysik and Bethe Center for Theoretical Physics,
Universit\"at Bonn,  Bonn 53115, Germany }

\begin{abstract}

   Based on the non-relativistic 
   QCD factorization scheme, we study the hadroproduction of the
bottomonium states $\Upsilon(5S)$ and $\Upsilon(6S)$.
We argue to search for them in the final states $\Upsilon(1S,2S,3S)\pi^+\pi^-$, which are
found to have anomalously large production rates at $\Upsilon(5S)$.
The enhanced rates for the dipionic transitions in the $\Upsilon(5S)$-energy region could, besides $\Upsilon(5S)$, be ascribed to $Y_b(10890)$, a state reported by the Belle collaboration, 
which may be interpreted as a tetraquark. The LHC/Tevatron measurements are capable
of making a case in favor of or against the existence of $Y_b(10890)$, as demonstrated here.
Dalitz analysis of the  $\Upsilon(1S,2S,3S)\pi^+\pi^-$ states from the $\Upsilon(5S)/Y_b(10890)$
decays also impacts directly on  the interpretation of the charged bottomonium-like states,
$Z_b(10600)$ and $Z_b(10650)$, discovered by Belle in these puzzling decays.

\end{abstract}

\pacs{13.85.Ni;14.40.Pq;14.40.Rt }
\maketitle

As a multi-scale system,  heavy-quarkonium states 
provide a unique laboratory to explore the interplay between perturbative
and nonperturbative effects of  QCD. Due to the 
non-relativistic nature,  these states allow the application of theoretical tools that can simplify and
constrain the analyses of nonperturbative effects.  The commonly-accepted method 
 is  the non-relativistic
QCD (NRQCD)~\cite{Bodwin:1994jh} which adopts a factorization ansatz to separate the short-distance
 and long-distance effects.  
Since the bottom quark is approximately three times heavier than the charm quark, it is expected that
 the expansion in $\alpha_s(\mu)$, where $\mu$ is a scale of $O(m_b)$, and 
 $v^2$,  with $v$ as the velocity of the heavy quark in  the hadron, which also is an  NRQCD 
expansion parameter,  converges much faster  for  the bottomonium states. 
Consequently,  great progress has been made  in the past years on the hadronic production of
$\Upsilon(1S,2S,3S)$~\cite{Brambilla:2010cs}.  On the experimental side the production rates and polarization
have been measured at the Tevatron~\cite{Acosta:2001gv,Abazov:2008aa,CDF:2011ag} and at the 
LHC~\cite{Aad:2012yna,Aaij:2013yaa,Chatrchyan:2013yna}. Theoretical attempts to explain these data 
have been independently performed by several groups with the inclusion of the next-to-leading
order QCD corrections~\cite{Campbell:2007ws,Artoisenet:2008fc,Gong:2010bk,Butenschoen:2010rq,Butenschoen:2012px,Wang:2012is,Gong:2008hk,Gong:2013qka}. 
 
Experimental and theoretical  studies  performed at hadron colliders have so far been
limited  to the $\Upsilon(1S,2S,3S)$ bound states,
since they all lie below the $B\bar{B}$ threshold and hence  have sizable leptonic branching
fractions. Above the $B\bar{B}$ threshold, however, the leptonic branching ratios of the higher
bottomonium states become very small, as a consequence of which these states have not been
seen so far in hadronic collisions. But, if the anomalously large decay widths of $O(1)$ MeV
in the final states $\Upsilon(1S,2S,3S) \pi^+\pi^-$, reported by Belle a few years
ago~\cite{Abe:2007tk,:2008pu}, are to be ascribed to the decays of the $\Upsilon(5S)$, then these final states are also promising for
the detection of the $\Upsilon(5S)$  in experiments at the Tevatron and the LHC~\cite{Ali:2011qi}. Arguing along similar lines,
the rescattering mechanism which enhances the dipionic partial widths in the $\Upsilon(5S)$ decays is 
also likely to yield similar enhancements in the rates for the corresponding transitions in the
$\Upsilon(6S)$ decays~\cite{Meng:2007tk}, which then could also be measurable in hadronic collisions. In this paper,
we derive the hadroproduction cross sections for $\Upsilon(5S)$ and $\Upsilon(6S)$ in $p\bar p(p)$ collisions
using  the NRQCD framework, supplemented by the subsequent decays into $\Upsilon(1S,2S,3S) \pi^+\pi^-$.

The enhanced rates for the dipionic transitions in the $\Upsilon(5S)$-energy region could,
however, also be ascribed to $Y_b(10890)$, a state reported by the Belle
collaboration~\cite{Abe:2007tk,:2008pu}, 
which is tentatively interpreted as a tetraquark~\cite{Ali:2009pi,Ali:2009es,Ali:2010pq,Ali:2011ug}.
In that case, one expects a smaller cross section
for the hadroproduction of $Y_b(10890)$ than for a genuine $b\bar{b}$ bound state. At the same time, 
as there are no tetraquark states expected to lie in the $\Upsilon(6S)$ region,
there would be no plausible grounds to expect a measurable yield in the
$(\Upsilon(1S,2S,3S)\to \mu^+\mu^-)\pi^+\pi^-$ final states from the decays of $\Upsilon(6S)$.  
 Since exotic states
in the charm sector have been successfully searched for in the $(J/\psi, \psi^\prime) \pi^+\pi^-$ final
states not only at the $e^+e^-$ colliders, but also  in hadroproduction in experiments at the Tevatron~\cite{Abulencia:2006ma} and the 
LHC~\cite{Aaij:2013zoa,Chatrchyan:2013cld}, the proposed measurements at hadron colliders
in the final states $(\Upsilon(1S,2S,3S)\to \mu^+\mu^-)\pi^+\pi^-$ could open new avenues in the search
and discovery of the exotic four quark states in the bottom  sector. 
 In particular, there exist three candidates up to date, namely the states
 labeled $Y_b(10890)$, $Z_b(10610)$ and $Z_b(10650)$, with the last two observed by Belle last
 year~\cite{Belle:2011aa}.
If the exotic state $Y_b(10890)$ is not confirmed, we have nonetheless demonstrated a new way to explore
the bottomonia  above the $B\bar{B}$ threshold, which would supplement the study of
$\Upsilon(1S,2S,3S)$ in hadronic collisions. First steps in that direction are recently reported by the CMS collaboration~\cite{CMSconstraint}.

The cross section for  the hadroproduction process $ p\bar p(p)\to\Upsilon+ X$ (we will leave $X$ implicit in the following) is given by
\begin{eqnarray}
\hspace{-10pt}\sigma_N( p\bar p(p)\to \Upsilon+X)\hspace{-2pt}&=&\hspace{-2pt} \int dx_1 dx_2 \sum_{i,j} f_i(x_1) f_j(x_2) \nonumber\\
&&\times  \hat \sigma (ij\to  \langle \bar bb\rangle_N +X) \langle O[N]\rangle, 
\end{eqnarray}
where $i,j$ denotes a generic parton inside a proton/antiproton, and $f_a(x_1), f_b(x_2)$ are the parton distribution
 functions (PDFs), 
which depend on the fractional momenta $x_i (i=1,2)$ (an additional scale-dependence is suppressed here), and $\Upsilon$ denotes a generic bottomonium state above $B\bar B$ threshold for which we consider $\Upsilon(5S)$ and
 $\Upsilon(6S)$ in this paper.  
We   adopt the CTEQ 6 PDFs~\cite{Nadolsky:2008zw} in our numerical calculations.
$ \langle O[N]\rangle$ are the long-distance matrix elements (LDMEs).
$N$ denotes all the quantum numbers of the $b \bar b$ pair, which we label in the form $^{2S+1} L_J^{c}$ (color $c$, spin $S$, angular momentum $L$, and total angular momentum $J$), and 
$\hat \sigma$ denotes the partonic cross section.

\begin{figure}   
\includegraphics[width=0.4\textwidth]{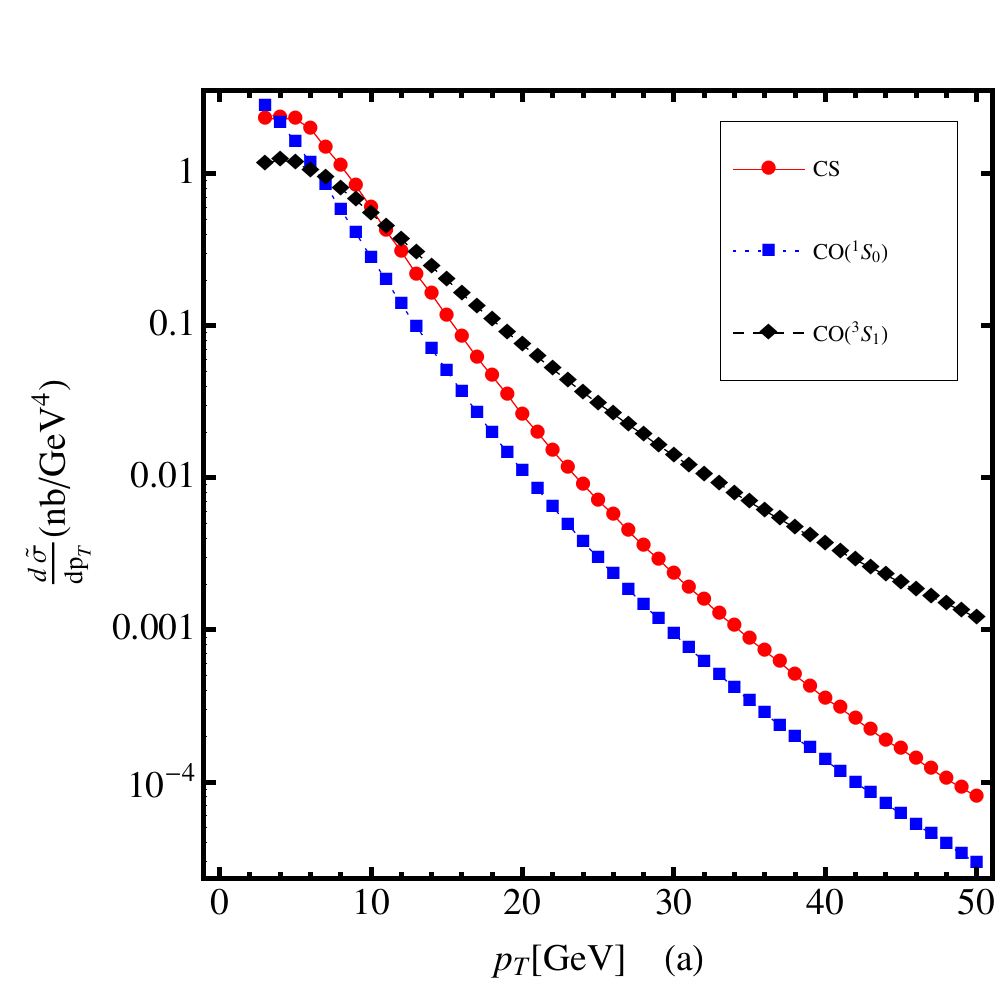} 
\includegraphics[width=0.4\textwidth]{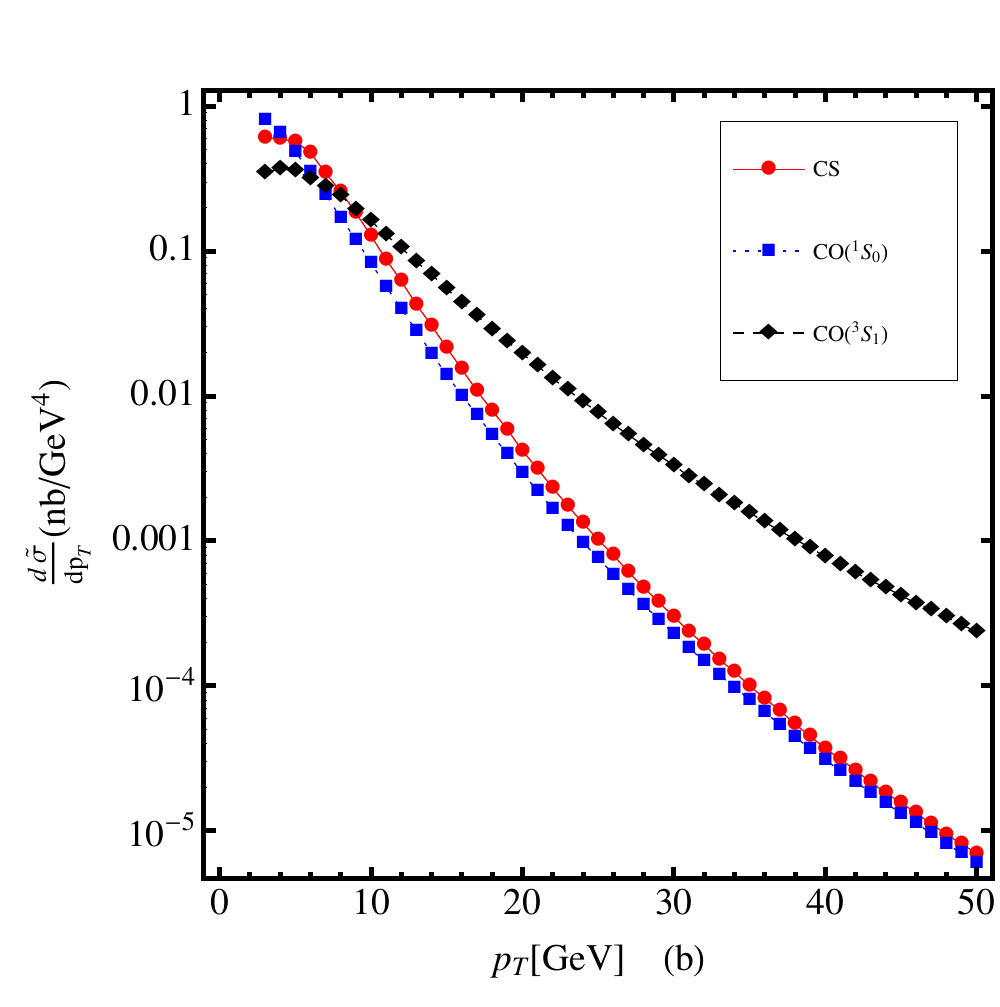}
\vspace{-5pt}
\caption{\label{fig_ptPlot} 
Individual contributions ($^3S_1^1$ solid, $^3S_1^8$ dashed, $^1S_0^8$ dotted, CO contributions are multiplied by $10^{-2}$) for the normalized transverse momentum  distributions $d\tilde\sigma/dp_t$ (explained in the text)  
for the process $pp\to \Upsilon(5S)$ at the LHC with $\sqrt s= 7$ TeV
for $|y| < 2.5$ (a) and $2 < |y| < 4.5$ (b)
 (the corresponding curves for $\Upsilon(6S)$ are almost identical
on a logarithmic plot).  
The $p_t$ integrated values are given in Tab.~\ref{tab:integratedCrossSection}.
 }
\end{figure}

The normalized cross sections, in which the LDMEs are factored out
are defined by  $ \tilde \sigma_N
 \equiv
 \sigma_N/\langle O[N]\rangle$.
The transverse momentum distribution is then  given by
\begin{equation}\label{eq:tcrosssec}
\hspace{-1pt}\frac{d\sigma_N}{d p_t}
\hspace{-2pt}=\hspace{-2pt}
\hspace{-2pt}\sum\limits_{i,j}
\int\hspace{-2pt}
J dx_1 dy  f_{i}(x_1,\mu_f)f_{j}(x_2,\mu_f)\frac{d \hat\sigma_N}{d t}\langle O[N]\rangle,\hspace{-2pt}
\end{equation}
where $y$ is the rapidity of $\Upsilon$, $p_t$ is the transverse momentum and  $J$ is the Jacobian  factor.
\begin{table}[t] 
\caption{ Integrated normalized cross sections $\tilde\sigma_N$, shown in Fig.~\ref{fig_ptPlot} 
(in units of nb$/$GeV$^3$, CO channels are multiplied by $10^{-2}$)  for the processes
$ p \bar{p}(p) \to  \Upsilon(5S,6S)$,
assuming a
transverse momentum range $3~{\rm GeV} <p_t<50~{\rm GeV}$.  
The 
rapidity range $|y| < 2.5 $  has been  assumed for the
Tevatron  experiments (CDF and D0)  at 1.96 TeV and for the LHC experiments (ATLAS and CMS) at 7, 8 and 14 TeV;  
 the 
 rapidity range $2.0<y<4.5$ is used for the LHCb. 
 }\label{tab:integratedCrossSection}
 \begin{tabular}{lrrrrrr}
 \hline\hline
 &
 &
 $\Upsilon(5S)$
 &
 &
 & 
 $\Upsilon(6S)$
 &
 \\
& $^3 S^1_1$ & $^3 S^8_1$ & $^1 S^8_0$
 & $^3 S^1_1$ & $^3 S^8_1$ & $^1 S^8_0$
 \\\hline
Tevatron &
2.72 & 1.73 & 1.75 & 2.54 & 1.66 & 1.60 \\  
LHC\phantom{b1} 7 & 
13.25 & 9.49 & 9.06 & 12.44 & 9.16 & 8.32 \\  
 LHCb\phantom{1} 7   & 
3.13 & 2.78 & 2.65 & 2.93 & 2.67 & 2.43 \\  
LHC\phantom{b1} 8  &  
15.35 & 11.15 & 10.57 & 14.41 & 10.75 & 9.73 \\  
 LHCb\phantom{1} 8   & 
3.80 & 3.35 & 3.17 & 3.56 & 3.22 & 2.92 \\  
LHC\phantom{b} 14&
27.62 & 21.15 & 18.76 & 25.98 & 20.48 & 17.30 \\ 
LHCb 14  &
 7.99 & 6.91 & 6.45 & 7.50 & 6.67 & 5.92\\
 \hline\hline
 \end{tabular} 
 \vspace{-10pt}
\end{table}

The leading-order partonic processes for the S-wave configurations are:
\begin{align} 
g(p_1)g(p_2)
&\to
\Upsilon[^3S_1^{1}](p_3)+g(p_4),
\nonumber\\
g(p_1)g(p_2)
&\to
\Upsilon[^1S_0^{8},\; ^3S_1^{8}](p_3)+g(p_4), 
\nonumber\\
g(p_1)q(p_2)
&\to
\Upsilon[^1S_0^{8},\; ^3S_1^{8}](p_3)+q(p_4), 
\nonumber\\
q(p_1)\bar q(p_2)
&\to
\Upsilon[^1S_0^{8},\; ^3S_1^{8}](p_3)+g(p_4). 
\end{align}
These differential partonic cross sections, which are needed in Eq.~\eqref{eq:tcrosssec} have been calculated  in fixed-order perturbation theory in the literature.
 For the color singlet (CS), one has~(see for instance Ref.~\cite{Gong:2008hk}):
\begin{eqnarray}
\frac{d \hat \sigma}{d \hat t}
=
\frac{5 \pi^2 \alpha_s^3 
[\hat s^2(\hat s-1)^2 +\hat t^2(\hat t -1)^2+\hat u^2(\hat u-1)^2 ]}{216 m_b^5 \hat s^2(\hat s-1)^2(\hat t-1)^2(\hat u-1)^2}.
\end{eqnarray}
The normalized  Mandelstam variables are defined as
\begin{align}
\hat s&=\frac{(p_1+p_2)^2}{4 m_b^2} ,\;\;
\hat t=\frac{(p_1-p_3)^2}{4 m_b^2} ,\;\;
\hat u=\frac{(p_1-p_4)^2}{4 m_b^2},
\end{align}
with $ m_b\simeq 4.75$ GeV. 
The factorization scale $\mu_f$ is chosen as $\mu_f=\sqrt{4m_b^2+ p_t^2}$. 
The partonic cross sections for the color octet (CO) have been calculated in Refs.~\cite{Gong:2010bk,Cho:1995ce}.
The $K$-factor for the CS contribution has been calculated for the process $p\bar p(p)\to \Upsilon(1S)$
in~\cite{Gong:2008hk}, which we have employed for the numerical calculations presented here. 
It is assumed, that the $K$-factor is not sensitive to $\sqrt{s}$. The CO contributions are taken at LO, since the
NLO corrections, which have also been calculated for $p\bar p(p)\to \Upsilon(1S)$,  are small~\cite{Gong:2010bk}.

Using these inputs, we show  the  
transverse momentum  distributions ${d\tilde\sigma}/{dp_t}$ in Fig.~\ref{fig_ptPlot} for the processes
$ p p \to  \Upsilon(5S)$ at   the LHC with $\sqrt s=7 $ TeV in the transverse momentum range
 $3~{\rm GeV} <p_t<50~{\rm GeV}$, where  $\log(p_t/m_{\Upsilon(5S)})$ is not large enough to necessitate
the resummation of the logarithms~\cite{Kang:2011mg,Sun:2012vc,Ma:2012hh}. 
The integrated normalized cross sections $\tilde \sigma_N$ 
are given in in Tab.~\ref{tab:integratedCrossSection}.

For the long-distance part we need nonperturbative input. The CS-LDMEs are given by the radial wave function at
the origin and can be extracted from the partial $e^+e^-$ widths via the Van-Royen Weisskopf formula.
Using the Particle Data Group values~\cite{Beringer:1900zz} for the leptonic partial widths as input, and
$ m_{\Upsilon(5S)}=10876 $~MeV, $m_{\Upsilon(6S)}=11019 $~MeV, we find at NLO 
$|R(0)|^2_{\Upsilon(5S)} =  2.37$~GeV$^3$ and   
$|R(0)|^2_{\Upsilon(6S)} =  1.02$~GeV$^3$.
The radial  wave function at origin $R(0)$  is related to the LDME via $ <O^H\;^3 S^1_1> =3|R(0)|^2/(4\pi)$.

The CO-LDMEs can  only be extracted  from the experimental data on differential distributions.
This has been done for the  $\Upsilon(1S, 2S, 3S)$  states by fitting the data on  $\Upsilon(1S, 2S, 3S)\to \mu^+\mu^-$.
We do not have the corresponding nonperturbative input for $\Upsilon(5S, 6S)$ at the current stage.
Once the $p_t$-distributions in these states have been measured, the CO matrix elements can be extracted from an
NRQCD-based analysis of the data. 

\begin{figure}   
\includegraphics[width=0.22\textwidth]{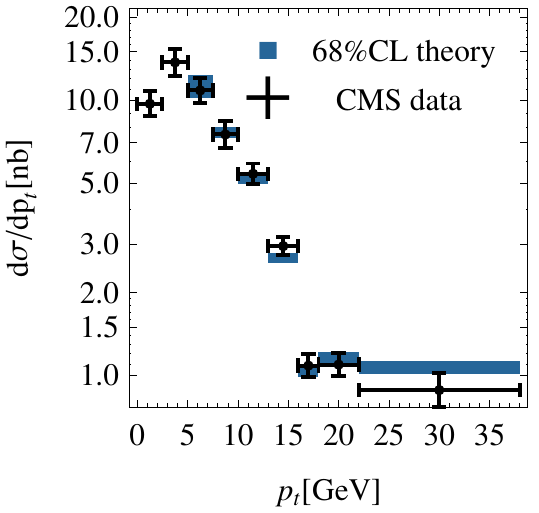}
\raisebox{-2pt}{\includegraphics[width=0.21\textwidth]{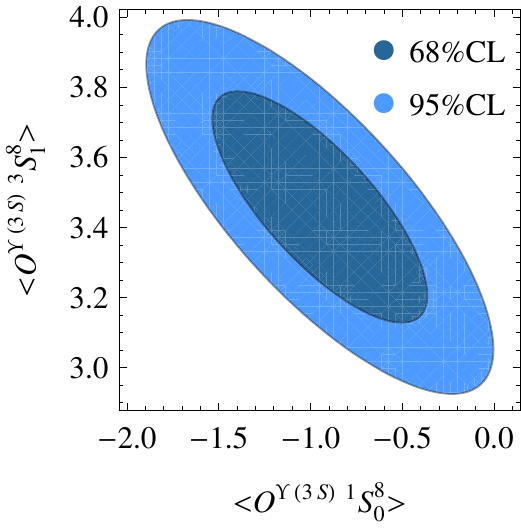} }
\vspace{-5pt}
\caption{\label{fig_estfrom3S} 
 Estimates of the LDMEs for $\Upsilon(3S)$ by fitting data on $\Upsilon(3S)\to\mu^+\mu^-$ by the  CMS collaboration~\cite{Chatrchyan:2013yna} with $p_t>3$~GeV. The fit is shown to the left, while the error correlation of the parameters is shown to the right.
 }
\end{figure}

For this work we attempt an estimate for $\Upsilon(5S)$ and $\Upsilon(6S)$ 
by adopting the full range from no CO contribution (lower bound) to the LDMEs estimated from $\Upsilon(3S)$ (upper bound).
Generically it is believed that the contribution of the CO-LDMEs in the hadroproduction of the $\Upsilon(nS)$ decreases as the principal number number $n$ increases.  Hence, our predicted range  can be viewed as  conservative.
We extract the values from data on $pp\to \Upsilon(3S)$ by the CMS collaboration~\cite{Chatrchyan:2013yna} for $p_t>3$~GeV. The advantage of using the estimates from $\Upsilon(3S)$ is the negligibility of feeddown contributions of higher lying states, which is for example present in $\Upsilon(1S,2S)$ and  makes the extraction somewhat more biased.
We find 
$<O^H\;^1 S^8_0>= (-0.95\pm 0.38)10^{-2}\rm{GeV}^3$   and
$<O^H\;^3 S^8_1>= (3.46\pm 0.21)10^{-2}\rm{GeV}^3$
and obtain good agreement with the data with
$\chi^2/\rm{d.o.f.}  =  4.3/5$, depicted in Fig.~\ref{fig_estfrom3S} (left-hand).
Our findings are in agreement with~\cite{Gong:2013qka}.
There is a small discrepancy of $1.2\sigma$ with $<O^H\;^3 S^8_1>=(2.71 \pm 0.13)10^{-2}\rm{GeV}^3$, which may be due to the fact, that in~\cite{Gong:2013qka} also $P$-wave contributions are considered, which can account for the smaller value.
Refitting the data has several advantages; we can account for large error correlations, shown in Fig.~\ref{fig_estfrom3S} (right-hand), of the LDMEs and get a consistent framework from the extraction to the final theory prediction. 
Certainly, the measurement of the $\Upsilon(5S)$ transitions will also strongly improve the estimates for the $\Upsilon(6S)$ by getting further information about the LDMEs of the higher lying states. 


For the exclusive production processes $p\bar p(p)\to \Upsilon(5S,6S)\to (\Upsilon(1S,2S,3S)\to\mu^+\mu^-) \pi^+\pi^-$, 
we combine the results from the cross sections discussed earlier and the branching ratios for the $\Upsilon(nS)$ decays,
which are listed in Tab.~\ref{tab:upsinfo}. Note, that in this table, the branching fractions of $\Upsilon(5S) \to \Upsilon(1S,2S,3S)\pi^+\pi^-$ are taken from the experimental data, and thus no model-dependence is introduced. For the estimate for the branching ratio of $\Upsilon(6S)$, we  rely on  the rescattering model~\cite{Meng:2007tk} as  the Belle anomaly seen in the $\Upsilon(5S)$ decays can be  well explained. 
In this model, the $\Upsilon(6S)\to \Upsilon(1S,2S,3S)\pi^+\pi^-$ channels are also expected to have partial widths
of about $1$ MeV. Possible variations in the estimates of the off-mass-shell effects due to different parametrizations
are neglected, as we don't expect them to be large. 
Again we remain on the conservative side by adding the errors on the branching ratios linearly for the upper and lower bounds.
\begin{table}[t]
\caption{Branching ratios for the $\Upsilon(5S)$ and $\Upsilon(6S)$.
 All input values are taken from the PDG~\cite{Beringer:1900zz}, except for the  $\Upsilon(6S)$ entries,
which are estimated from the scattering model~\cite{Meng:2007tk}.
}
\label{tab:upsinfo}
\begin{center}
\begin{tabular}{ll}
\hline 
\hline 
${\cal B} (\Upsilon(5S)\to \Upsilon(1S) \pi^+\pi^-)$ & $(0.53\pm 0.06)\%$  \\
${\cal B} (\Upsilon(5S)\to \Upsilon(2S) \pi^+\pi^-)$ &$(0.78\pm 0.13)\%$    \\ 
${\cal B} (\Upsilon(5S)\to \Upsilon(3S) \pi^+\pi^-)$ &$(0.48
\pm 0.18)\%$ 
\\
\hline
${\cal B} (\Upsilon(6S)\to \Upsilon(1S)\pi^+\pi^-)$ & $\approx 0.4  \%$ 
\\
${\cal B} (\Upsilon(6S)\to \Upsilon(2S)\pi^+\pi^-)$ & $(0.4 - 1.2)  \%$ 
\\
${\cal B} (\Upsilon(6S)\to \Upsilon(3S)\pi^+\pi^-)$ & $(1.2-2.5) \%$ 
\\
\hline 
${\cal B}(\Upsilon(1S)\to \mu^+\mu^-)$ &$ (2.48\pm0.05)\%$ \\
${\cal B}(\Upsilon(2S)\to \mu^+\mu^-)$ &$ (1.93\pm0.17)\%$ \\ 
${\cal B}(\Upsilon(3S)\to \mu^+\mu^-)$ &$ (2.18\pm0.21)\%$ \\
\hline 
\hline
\end{tabular}
\end{center}
\vspace{-20pt}
\end{table}

\begin{table*}
\caption{Total cross sections for the processes
$p\bar p(p)\to \Upsilon(5S,6S)\to (\Upsilon(nS)\to \mu^+\mu^-) \pi^+\pi^-$
$(n=1,2,3)$ in pb
at the  Tevatron ($\sqrt{s}=1.96$ TeV) and the LHC 
($\sqrt{s}=$7, 8, 14 TeV), assuming 
the rapidity intervals described in 
Table~\ref{tab:integratedCrossSection}.
The error estimates are from the variation of 
the central values of the CO-LDMEs and the various decay branching ratios,
see text.}
\label{tab:fullresult}
\begin{tabular}{lccccccc}
\hline\hline
 &
 &
 $\Upsilon(5S)$
 &
 &
 & 
 $\Upsilon(6S)$
 &
 \\
& $n=1$ & $n=2$ & $n=3$
 & $n=1$ & $n=2$ & $n=3$\\\hline
Tevatron
 &[0.18,0.98]
 &[0.18,1.35]
 &[0.09,1.03]
 &[0.06,0.57]
 &[0.04,1.38]
 &[0.15,3.26]
\\
LHC  7
 &[0.86,5.26]
 &[0.86,6.74]
 &[0.44,5.56]
 &[0.29,3.13]
 &[0.21,7.57]
 &[0.72,17.9]
\\
LHCb 7
 &[0.20,1.48]
 &[0.20,1.89]
 &[0.10,1.56]
 &[0.07,0.89]
 &[0.05,2.16]
 &[0.17,5.13]
\\
LHC 8
 &[0.99,6.17]
 &[0.99,7.89]
 &[0.51,6.52]
 &[0.34,3.67]
 &[0.25,8.87]
 &[0.83,21.0]
\\
LHCb 8
 &[0.25,1.78]
 &[0.25,2.28]
 &[0.13,1.88]
 &[0.08,1.08]
 &[0.06,2.61]
 &[0.20,6.19]
\\
LHC 14
 &[1.79,11.7]
 &[1.79,14.9]
 &[0.92,12.3]
 &[0.61,7.02]
 &[0.45,17.0]
 &[1.50,40.2]
\\
LHCb 14
 &[0.52,3.70]
 &[0.52,4.74]
 &[0.27,3.91]
 &[0.18,2.25]
 &[0.13,5.43]
 &[0.43,12.9]
\\
 \hline\hline
\end{tabular} \end{table*}

The total cross sections for the processes
$p\bar p(p)\to \Upsilon(5S,6S)\to (\Upsilon(nS)\to \mu^+\mu^-) \pi^+\pi^-$
$(n=1,2,3)$ in pb
at the  Tevatron ($\sqrt{s}=1.96$ TeV) and the LHC 
($\sqrt{s}=$7, 8, 14~TeV), assuming 
the rapidity intervals described in 
Tab.~\ref{tab:integratedCrossSection} are given in Tab.~\ref{tab:fullresult}.
They are the principal results derived in this paper and deserve a number of comments.
In calculating the cross sections, we have included the
next-to-leading order contributions by rescaling the available results for the
process $p\bar p(p)\to \Upsilon(1S)$~\cite{Gong:2008hk}.
 Fixing the center-of-mass energy and the rapidity interval, the cross-sections for each of the six processes
$pp(\bar{p} \to \Upsilon(5S, 6S) \to (\Upsilon(nS) \to \mu^+\mu^-) \pi^+\pi^-$ $(n=1,2,3)$ are 
bounded in an interval spanning roughly one order of magnitude, mainly due to our current  ignorance of the
CO matrix elements for the  $\Upsilon(5S)$ and $\Upsilon(6S)$. 
 We note that the cross sections increase by an order of magnitude in going from the Tevatron ($p\bar{p}; \sqrt{s}=1.96$ TeV)
to the LHC ($pp$; $\sqrt{s}=14$ TeV), with the cross sections at the 8 TeV (with the currently highest luminosity of about
20 (fb)$^{-1}$) is typically of $O(1)$ pb for the ATLAS and the CMS experimental setups. 
Thus, the estimates presented here offer achievable targets for the future experimental searches.

Recently, the CMS  collaboration has presented the search  for a new bottomonium state, denoted as $X_b$,
decaying to $\Upsilon(1S)\pi^+\pi^-$, based on a data sample corresponding to an integrated luminosity of 20.7 ${\rm fb}^{-1}$ at $\sqrt{s} $= 8 TeV~\cite{CMSconstraint}.   No evidence is found for $X_b$, and the upper limit at a confidence level of 95\% on the production cross section of $X_b$ times the decay branching fraction of $X_b\to \Upsilon(1S)\pi^+\pi^-$ is set to be
\begin{eqnarray}
 \frac{\sigma(pp\to X_b\to \Upsilon(1S)\pi^+\pi^-)}{\sigma(pp\to \Upsilon(2S)\to \Upsilon(1S)\pi^+\pi^-)} < 0.02,
 \label{eq:CMSratio}
\end{eqnarray}
where the stated upper bound correspond to the $X_b$ mass around 10.876 GeV. 

Using the current experimental data on the $\sigma(pp\to \Upsilon(2S))$, we can convert the above ratio to absolute cross sections and compare with our results. Since the masses of the $\Upsilon(2S)$ and $X_b$ are close,  it may be a good approximation to assume that the   ratio given in Eq.~\eqref{eq:CMSratio} is insensitive to the kinematics cuts. Using the CMS result at  $\sqrt s=7$ TeV in Ref.~\cite{Chatrchyan:2013yna}:
\begin{eqnarray}
 \sigma (pp\to \Upsilon(2S)X ) {\cal B}(\Upsilon(2S)\to \mu^+\mu^-) 
   = 1.55~{\rm nb}
\end{eqnarray}
we get  
\begin{eqnarray}
 \!\!\sigma\! (pp\!\to\! X_b\!\to\!\Upsilon(1S) \pi\!^+\pi\!^-\!) {\cal B}(\Upsilon(1S)\!\to\! \mu\!^+\mu\!^-)  
  \!<\! 7.1~{\rm pb},
\end{eqnarray}
where the ranges 3 GeV $<p_t< $50 GeV and $|y|<2.4$ have been used. Identifying $X_b$ with the $\Upsilon(5S)$,  the above upper bound is 
larger than our predictions by a factor of $\mathcal{O}(10)$. 
 With increased luminosity, the next round of experiments at the
   LHC will reach the required experimental sensitivity. We also note
       that the current CMS analysis is based on a stiff cut on $p_t$
       ($p_t > 13.5$ GeV). Lowering this cut will help significantly in
       the discovery of $X_b$ ($\Upsilon(5S)$ or $Y_b$).


There are two competing scenarios, which can be explored in the future data analysis:

i) Experiments are able to establish the signals in the processes 
$p\bar p(p)\to \Upsilon(5S,6S)\to (\Upsilon(1S,2S,3S)\to \mu^+\mu^-) \pi^+\pi^-$, in approximate agreement with the
estimates presented here, based on the rescattering formalism. This would then extend the application of the NRQCD techniques to the yet unexplored
sectors $\Upsilon(5S)$ and $\Upsilon(6S)$ in hadroproduction. 
Note, that the estimates can be  easily adopted for any other scenario, 
which relies on the anomaly coming from $\Upsilon(5S)$.        In this case also, one expects the $\Upsilon(6S)$ to reflect the
       enhanced branching ratios to $\Upsilon(nS) \pi^+ \pi^-$, as seen
       in the decays of $\Upsilon(5S)$. Any such scenario will be tested
       as well.  Changing the estimated branching ratio, listed in Tab.~\ref{tab:upsinfo}, is  straight forward.

ii) Experiments  are able to establish only the process 
 $p\bar p(p)\to \Upsilon(5S)\to (\Upsilon(1S,2S,3S)\to \mu^+\mu^-) \pi^+\pi^-$, but not
$p\bar p(p)\to \Upsilon(6S)\to (\Upsilon(1S,2S,3S)\to \mu^+\mu^-) \pi^+\pi^-$, which, in our opinion,
would speak against the rescattering mechanism and strengthen the case of $Y_b(10890)$ as the source of
 the anomalous  dipion transitions~\cite{Ali:2009pi,Ali:2009es,Ali:2010pq,Ali:2011ug}. In this case, experiments
would provide a calibration of the cross section for the exotic hadron $Y_b(10890)$ -- certainly a valuable piece
of information in this unexplored QCD sector.

Once enough data are available, one could undertake a Dalitz analysis of the $\Upsilon(1S, 2S, 3S)\pi^+\pi^-$
final states to determine the origin of the charged tetraquarks states
$Z_b(10610)$ and $Z_b(10650)$, discovered  by the BELLE collaboration~\cite{Belle:2011aa} along
similar lines. In this regard, we wish to
point out that recently a charged four-quark state $Z_c(3900)$ has been discovered by the BESIII
collaboration~\cite{Ablikim:2013mio}, confirmed by Belle~\cite{Liu:2013dau}, in the decays
$Y(4260) \to Z_c(3900)^\pm \pi^\mp \to J/\psi \pi^+\pi^-$,
where $Y(4260)$ is an exotic $c\bar{c}$ state~\cite{Brambilla:2010cs}, possibly a
tetraquark~\cite{Ali:2011ug,Faccini:2013lda}. Another charged four-quark state $Z_c(4025)$ is found in  $e^+e^- \to (D^{*} \bar{D}^{*})^{\pm} \pi^\mp$ at $\sqrt{s}=4.26$\,GeV by the BESIII collaboration~\cite{Ablikim:2013emm}. 
 These observations  indirectly support  the interpretation that
$Z_b(10610)$ and $Z_b(10650)$ are likewise the decay products of the exotic state $Y_b(10890)$. 
We note that $Z_c(3900)$ is also found in the analysis based  on CLEO data~\cite{Xiao:2013iha}.   These charmonium-like states can   be accessed  at hadron colliders in the final state $J/\psi^{(')}\pi^+\pi^-$. 
 

Using the available  NRQCD results, we have explored the hadroproduction of 
bottomonium states  above the $B\bar B$ threshold  at the LHC and the Tevatron.
The large branching fractions for the decays $\Upsilon(5S) \to \Upsilon(1S,2S,3S)\pi^+\pi^-$, observed by
Belle~\cite{Abe:2007tk, :2008pu},  offer an opportunity to access the $\Upsilon(5S)$ in hadronic collisions.
Attributing the large
enhancement in the dipionic transitions at the $\Upsilon(5S)$ to the rescattering
phenomenon~\cite{Meng:2007tk}, very similar dipionic rates are expected for the  $\Upsilon(6S)$
decays, which we have also worked out and can easily be generalized for other scenarios. 
By the above computations, with results shown in Tab.~\ref{tab:fullresult}, we have shown that the experiments at the hadron colliders LHC  have in principle the sensitivity to detect
the bottomonium states $\Upsilon(5S)$ and  $\Upsilon(6S)$,
extending significantly their current experimental reach, and exploring thereby also the nature of  the exotic states $Y_b(10890)$, $Z_b(10600)$ and $Z_b(10650)$, discovered in $e^+e^-$ annihilation experiments. 
As the observed decay widths for $\Upsilon(5S) \to
       \Upsilon(nS) \pi^+ \pi^-$ show an anomalous enhancement by two
       orders of magnitude, and our estimates presented in Tab.~\ref{tab:fullresult}
       are uncertain by at most an order of magnitude, the rescattering
       mechanism as the source of this enhancement can be tested. Once
       the $p_t$-spectrum of the $\Upsilon(5S)$ is measured, the data
       will be precise enough to unambiguously pin down the nature of the
       resonance $\Upsilon(5S)$ or $Y_b$.
Our lower bounds presented in Tab.~\ref{tab:fullresult} puts the measurement of the $\Upsilon(5S)$ within reach of the next round of experiments at the LHC.

{\it Acknowledgments.}
W.W.  is  grateful to Bin Gong, Yu Jia,  Peng Sun and Jian-Xiong Wang for valuable discussions.  We thank Kai-Feng Chen for useful communication on the recent CMS results in Ref.~\cite{CMSconstraint}. 
The work of W.W.   is supported   by the DFG and the NSFC through funds provided to
the Sino-German CRC 110 ``Symmetries and the Emergence
of Structure in QCD".

\end{document}